\documentclass[review]{elsarticle}
\usepackage[utf8]{inputenc}
\usepackage{amssymb,amsfonts,amsmath}
\usepackage{physics}
\usepackage{xcolor}
\usepackage{yhmath}
\usepackage{float}
\usepackage{ulem}
\usepackage{epstopdf}
\biboptions{numbers,sort&compress}

\title{Kenfack \--- $\dot{\text{Z}}$yczkowski indicator of nonclassicality  for two non-equivalent representations of  Wigner function of qutrit}

\author[1,2]{Vahagn Abgaryan}

\author[1,3,4]{Arsen~Khvedelidze}

\author[1]{Astghik~Torosyan}

\address[1]{Laboratory of Information Technologies, Joint Institute for Nuclear Research, Dubna,  Russia}
\address[2]{Peoples’ Friendship University of Russia (RUDN University)}
\address[3]{Institute of Quantum Physics and Engineering Technologies, Georgian Technical University, Tbilisi, Georgia}
\address[4]{A.Razmadze Mathematical Institute, Iv.Javakhishvili Tbilisi State University, Tbilisi, Georgia }

\begin{document}

\begin{abstract}

Following Kenfack and $\dot{\text{Z}}$yczkowski, we consider the indicator of nonclassicality of quantum states for $N\--$level systems defined via the integral of the absolute value of 
the Wigner function. For these systems, remaining  in the framework of Stratonovich-Weyl correspondence, one can construct a  whole family of representations of the Wigner functions defined over the continuous phase-space and 
characterized by a set of $(N-2)$ moduli parameters.
It is shown that the nonclassicality indicator, being  invariant under the $SU(N)$ transformations of states, turns to be sensitive to the representation of the Wigner function.
We analyse this representation dependence computing the  Kenfack-$\dot{\text{Z}}$yczkowski indicators for pure and mixed states of a 3-level system using a generic and two degenerate Stratonovich-Weyl kernels respectively. 
Our calculations reveal three classes  of states: 
the ``absolutely classical/quantum'' states, which have zero and non-vanishing indicator for all values of the moduli parameters correspondingly,  and the ``relatively quantum-classical'' states whose classicality/quantumness is  susceptible to a representation of the Wigner function. 
Herewith, all pure states of qutrit belong to the ``absolutely quantum'' states.

\end{abstract}


\maketitle


\section{Introduction}

{\noindent $\bullet$\textbf{\, Negativity of Wigner Function\, }$\bullet$}
The Wigner function is a famous member of a peculiar  class of distributions,  the so-called  
\textit{quasiprobability distributions},  which have the prefix ``quasi'' in their names  because they do not  conform to the basic principle of true statistical distributions of being non-negative. This anomaly  is archetypal for all quantum systems: ``continuous'' and ``discrete''. 

For the first type of systems whose states are represented by density matrices $\varrho$ acting on the space of square-integrable functions $L^2(\mathbb{R})$ the Wigner function 
is defined over a 2-dimensional phase space with canonical coordinates $(x, p) \in \mathbb{R}^{2}\,$:
\begin{equation}
    \label{eq:CWF2}
    W_\varrho(x,p)=\frac{1}{\pi\hbar}\int_{-\infty}^{\infty} \mathrm{d}y\,\langle x+y|\,\varrho\,|x-y \rangle 
    e^{-2\imath py/\hbar}\,,
\end{equation}
with the well known bounds \cite{Wigner1932,HilleryEtAl1984}:
\begin{equation}
\label{eq:CWF2Bounds}
    -\frac{1}{\pi\hbar} \leq W(x,p) \leq \frac{1}{\pi\hbar}\,.
\end{equation}
Moreover, for the Wigner function defined on $2n$-dimensional phase space, the integrals over a domain $D
\in \mathbb{R}^{2n}\,$ are bounded by its volume
\cite{BrackenDoebnerWood1999}:
\begin{equation}
\label{eq:VolumeBoundary}
   -\frac{1}{{(\pi\hbar)}^n}\,
   {\mbox{Volume}(D)}
     \leq \int_D\,[\mathrm{d}x\mathrm{d}p]_N W(x,p) \leq
     \frac{1}{{(\pi\hbar)}^n}\,
   {\mbox{Volume}(D)}\,.
\end{equation}

Similarly, for the Wigner functions associated to a ``discrete'' $N$\--level quantum system, whose Hilbert space is $\mathcal{H}=\mathbb{C}^N$\,,
the analog of the bounds (\ref{eq:CWF2Bounds}) and (\ref{eq:VolumeBoundary}) exists.  Particularly,  in the framework of the  Weyl-Stratonovich  formalism 
(see, e.g., \cite{Stratonovich1957,BrifMann1998,RoweSandersGuise,Luis2008,KlimovedeGuise1,KlimovedeGuise2,TilmaEverittSamsonMunroNemoto2016,AKh2018} and references therein), it was shown in \cite{AKhT2019} that the Wigner quasiprobability distributions  $W_\varrho(\Omega_N)$ defined  over the phase-space $\Omega_N$,  obey  inequalities: 
\begin{equation}
\label{eq:WFNBounds}
  \sum_{i=1}^{N}\pi_ir_{N-i+1}\, \leq\,  W_\varrho(\Omega_N)\, \leq \sum_{i=1}^{N}\pi_ir_i\,, 
 \end{equation}
where $r_1, r_2. \dots, r_N$  are eigenvalues of  a mixed state $\varrho$ and $\pi_1, \pi_2, \dots \pi_N$ denote the eigenvalues of the Stratonovich-Weyl kernel with each set arranged in a decreasing order.

The bounds (\ref{eq:CWF2Bounds})-(\ref{eq:WFNBounds}),
being functions of  states and the Stratonovich-Weyl kernels eigenvalues, are exact in a sense that they are attainable at certain points of the phase space.  Particularly,  the subset of phase space where the Wigner function acquires negative values might be not empty. 
\footnote{
Definitely, there are states such that the bounds ~(\ref{eq:CWF2Bounds}) and ~(\ref{eq:VolumeBoundary}) are not optimal and the lower bound of the Wigner function for certain states can be positive. However, in  view of the well-known Hudson's theorem  \cite{Hudson1974}, a positive definiteness occurs only for  very special classes of states, e.g.,  a Gaussian wave function is the only pure state corresponding  to a positive Wigner function.
}

{\noindent $\bullet$\textbf{\, Wigner Functions Negativity vs. Classicality\, }$\bullet$} Faced with the negativity of probability distribution, Wigner in his 1932 paper \cite{Wigner1932} wrote:
``But of course this must not hinder the use of it in calculations as an auxiliary function which obeys many relations we would expect from such a probability''. This guideline turned out to be  foresighted. Almost a century of history of the method of quasiprobability distributions gave us an effective tool for analyzing quantum phenomena in variety of research areas including  quantum optics \cite{Hillery1987,LvovskyRaymer}, quantum theory of information and communications \cite{Ferrie2011,VeitchFerrieGrossEmerson2012}.
It turns out that during this time  a drastic metamorphosis happened in perception  of the negative distributions: from  an ``auxiliary'' probability function  up to proclamation of quasidistributions as a basic ingredient  in quantifying  the degree of quantumness (cf. recent discussions in
\cite{FerrieEmerson2009,FerrieMorrisEmerson2010,Ferrie2011,SperlingWalmsley2018}).

Relations between nonclassicality and Wigner function negativity become highly  intricate taking into consideration
existence  of infinitely many quasiprobability distributions  for a given quantum state.   
Nowadays this   aspect has drawn a wide attention especially in connection with   
a special class of the so-called discrete Wigner functions
\cite{Wooters1987,GibbonsHoffmanWootters2004}.
Particularly, very important findings has been done by R.W. Spekkens  about the interplay between negativity and nonclassicality.  It was demonstrated that the negativity is ``neither  a necessary nor a sufficient condition for the failure of classical explanation'' \cite{Spekkens2008}. Moreover, it was proved that for the discrete Wigner functions the  negativity is equivalent to the quantum contextuality and the  role of the contextualization of negativity of the Wigner function in a resource theoretical framework of non-Gaussianity was intensively  discussed
(see e.g., discussions in 
\cite{Spekkens2008,Delfosse1OkayVegaBrowneRaussendorf2017,RaussendorfBrowneDelfosseCVega2017,TakagiZhuang2018,AlbarelliGenoniParisFerraro2018,Ming}).
Below,  having in mind these results known for the discrete Wigner functions,  we are going to  discuss similar issues for the Wigner quasidistributions defined on a continuous phase-space. More precisely,  an analysis  of how the variety of the Wigner functions  representations  affects  a special measure of nonclassicality  will be given.

We focus here mainly on the three level quantum systems since the smallest system revealing contextuality is the qutrit \cite{KlyachkoCan, KurzynskiKaszlikowski1,ShrutiDograKavitaDoraiArvind}.

{\noindent $\bullet$\textbf{\, Degree of Negativity as Measure of Quantumness\, }$\bullet$}
Nonclassicality measures based on the  violation of the Wigner function semi-positivity can be divided into two different types. If one takes states with positive Wigner functions as the reference ''classical states''  then the measures of nonclassicality are based either on the distance from the set of ``classical states'' \cite{Hillery1987,DodonovEtAl2000,MarianEtAl2002}  or on the volume of a phase space region where the Wigner function is negative \cite{KenfackZyczkowski2004,AKhT2019}.
In the present note, following the second approach,  we will make use of the volume indicator of nonclassicality introduced by A. Kenfack and  K.$\dot{\text{Z}}$yczkowski \cite{KenfackZyczkowski2004} for an $N$-dimensional system admitting a Wigner function  defined over compact continuous phase space $\Omega_N$:
\begin{equation}
\label{eq:KZ}
    \delta(\varrho)=\int_{\Omega_N} \mathrm{d} \, \Omega_N\,\big|\,W_{\varrho}(\Omega_N)\,\big|-1\,.
\end{equation}

In definition (\ref{eq:KZ}), the notation  $\big|\,\cdot\,\big|$ stands for the absolute value (modulus) of the Wigner function. 
Hereinafter,
 the function $\delta(\varrho)$ will be termed as the 
KZ\--indicator.

In the next sections, the results of calculation of the   KZ\--indicators of nonclassicality (\ref{eq:KZ}) for two- and three-level systems,  qubits and qutrits respectively, will be given.
In the case of qutrit we will analyze a  functional dependence of the KZ\--indicator on the choice of representation of the Wigner function. Our computations are based on the description of the admissible representations of the Wigner function using the set of the so-called moduli parameters introduced in \cite{AKh2018}.
The analyses  shows that there is a special subset of ``absolutely classical'' states, such that their KZ\--indicator is vanishing independently of the Wigner function representation.
There is also a class of ``absolutely quantum'' states whose KZ\--indicator is non-zero for all values of  the Wigner function moduli parameter. Particularly, we will show that all pure states of  qutrit belong to the class of ``absolutely quantum'' states.

There are also ``relatively quantum-classical'' states whose classicality/quantumness depends on the representation of the Wigner function.

\section{Generalities on Wigner function of N-level system}

In this section, following presentations of
\cite{Stratonovich1957,BrifMann1998, AKh2018,AKhT2018},  we collect all necessary notations and definitions from  the  
Stratonovich-Weyl approach to the Wigner quasiprobability distribution  of a finite-dimensional system.

\noindent{$\bullet$ {\bf The Stratonovich-Weyl principles} $\bullet$}
Consider an  
$N$\--level quantum system in  a mixed state  characterized by a density matrix  $\varrho\,.$ 
Its expansion over the Hermitian basis  $\boldsymbol{\lambda} = \{\lambda_1\,,\cdots\,, \lambda_{N^2-1}\,\},$   
of $\mathfrak{su}(N)$ algebra
with the orthonormality conditions,  $\mbox{tr}(\lambda_\mu
\lambda_\nu) = 2\,\delta_{\mu\nu},$ 
reads
\begin{equation}
\label{eq:DM}
    \varrho = \frac{1}{N} \,\mathbb{I}_N+\sqrt{\frac{N-1}{2N}} \left(\boldsymbol{\xi},\boldsymbol{\lambda}\right)\,,
\end{equation}
where $\boldsymbol{\xi}$ is 
$(N^2-1)$-dimensional Bloch vector. 

The Wigner distribution   
$W_{\varrho}(\Omega_N)$ of an $N$-dimensional quantum system as a function on  symplectic  space $\Omega_N$ is defined  by pairing of a density matrix $\varrho$ and   the Stratonovich-Weyl kernel $\Delta(\Omega_N)\,,$
\begin{equation}
\label{eq:WF}
    W_{\varrho}(\Omega_N) = {\mbox{tr}(\varrho \, \Delta(\Omega_N))}\,. 
\end{equation}

The kernel $\Delta(\Omega_N)$ in (\ref{eq:WF}) obeys the following set of postulates, known under the name of  Stratonovich-Weyl correspondence \cite{Stratonovich1957,BrifMann1998}:
\begin{enumerate}
\item {\bf Reconstruction}: a state $\varrho$ is reconstructed from the Wigner function (\ref{eq:WF}) via the integral over a phase space: 
\begin{equation}\label{eq:DMWigner}
\varrho =\int_{\Omega_N} \mathrm{d}\Omega_N\, \Delta(\Omega_N) W_\varrho(\Omega_N) \,;
\end{equation}	
\item {\bf Hermicity}: 
\(\qquad
\Delta(\Omega_N)= \Delta(\Omega_N)^\dagger\,;
\)
\item {\bf Finite Norm}: a state norm is given by the integral of the Wigner distribution: 
\begin{equation}
\mbox{tr}[ \varrho ]= \int_{\Omega_N} 
\mathrm{d}\Omega_N W_\varrho(\Omega_N)\,, 
\qquad
\int_{\Omega_N} \mathrm{d}\Omega_N\,\Delta(\Omega_N) = 1\,;
\end{equation}
\item {\bf Covariance}: the adjoint unitary transformations  $U(\alpha) \in SU(N)\,$ of  a density matrix results  in the kernel change:
\begin{equation}
\label{eq:covSW}
\Delta^\prime(\Omega_N) =U(\alpha)^\dagger\Delta(\Omega_N)U(\alpha)\,, 
\end{equation}
via  symplectic transformations $ T_\alpha \in 
\mathrm{Sp}(d_N)$ on $\Omega_N\,:$ 
\footnote{
The rule  (\ref{eq:ztransf}) is an analogue of the well-known transformations generated by the metaplectic group $\mathrm{Mp}(n)$ of operators acting on $L^{n}(\mathbb{R})\,,$
cf. \cite{MdeGosson2006}.}
\begin{equation}
\label{eq:ztransf}
\Delta^\prime(\boldsymbol{z})=
    \Delta
    (\boldsymbol{z}^\prime)
    \,,\qquad 
 \boldsymbol{z}^\prime =
 T_{\alpha} \boldsymbol{z}\,. 
\end{equation}
Here  $\boldsymbol{z}=\{z_1, z_2,\dots\, z_{d_N}\} \in \Omega_N\,$ denote $d_N$ symplectic coordinates.
\end{enumerate}

As it was shown  in  \cite{AKh2018}, the above axioms are fulfilled if the Hermitian kernel $\Delta(\Omega_N)\,$ in (\ref{eq:WF}) 
satisfies  the following set of algebraic ``master equations'':
\begin{equation}
\label{eq:SWMaster}
    {\mbox{tr}(\Delta(\Omega_N))=1}\,, \quad 
    {\mbox{tr}(\Delta(\Omega_N)^2)=N}\,.
\end{equation}
These   equations determine the spectrum of Stratonovich-Weyl kernel $\mbox{\bf spec}(\Delta_N)=\{\pi_1(\boldsymbol{\nu}), \pi_2(\boldsymbol{\nu}), \dots, 
\pi_N(\boldsymbol{\nu}) \}\,$ non-uniquely and thereby cause the existence of the variety of representations for the Wigner functions. 

The corresponding moduli space of solutions to master equations (\ref{eq:SWMaster}) represents  a spherical polyhedron   on  $(N-2)\--$dimensional sphere  $\mathbb{S}_{N-2}(1)\,$ of radius one.
We denote the coordinates on moduli space by   $\boldsymbol{\nu}=\left(\nu_{1},\,\cdots,\,\nu_{N-2}\right)$ and hereafter will  point to the corresponding functional dependence of  the Stratonovich-Weyl kernel  and the Wigner function explicitly. 
See more on the moduli space of the Stratonovich-Weyl kernel in~\cite{AKhT2018}.

The phase space $\Omega_N$ is determined by the symmetries of  the  Stratonovich-Weyl kernel.  Assuming that the Stratonovich-Weyl kernel has a spectrum with the algebraic multiplicities  $\boldsymbol{k}=\{k_1, k_2, \dots, k_s\}$ 
then the   phase-space  can be identified with a complex flag variety 
$\Omega_{N,\boldsymbol{k}}
\simeq
\mathbb{F}^N_{\boldsymbol{k}}=
{U(N)} / {H}_{\boldsymbol{k}}\,, 
$
where the isotropy group of the  Stratonovich-Weyl kernel  $H_{\boldsymbol{k}}\in U(N)$ is of the form
\footnote{
The volume form on $\Omega_{N,\boldsymbol{k}}$ is given  by the 
bi-invariant normalised Haar measure $d\mu_{SU(N)}$ on $SU(N)$  group:
\(
d \Omega_{N,\boldsymbol{k}}=N\,\mbox{Vol}(H_{\boldsymbol{k}})\,{d\mu_{SU(N)}}/{d\mu_{H_{\boldsymbol{k}}}}\,,
\)
where $d\mu_{H_{\boldsymbol{k}}}$ is the induced measure on the isotropy group $H_{\boldsymbol{k}}$.}
\(
H_{\boldsymbol{k}}={U(k_1)\times U(k_2) \times \dots \times U(k_{s+1})}\,.
\)

Finalising this section, we give the expression for the Wigner function of an $N\--$ dimensional quantum system in terms of the Bloch vector $\boldsymbol{\xi}$ and a unit $(N^2-1)$\--dimensional vector $\boldsymbol{n}$ characterizing representative Stratonovich-Weyl kernel.
Using (\ref{eq:DM}) and the SVD decomposition 
of  the Stratonovich-Weyl kernel $\Delta(\Omega_N|\boldsymbol{\nu})$  :
\begin{equation}
\label{eq:SWk}
    \Delta(\Omega_N|\boldsymbol{\nu})=\frac{1}{N}U(\Omega_N)\left(\mathbb{I}_N+\kappa\sum_{\lambda_s\in K }\mu_s(\boldsymbol{\nu})\lambda_s\right)U(\Omega_N)^\dagger\,.
\end{equation}
In (\ref{eq:SWk}) $\kappa=\sqrt{{N(N^2-1)}/{2}}$\, is the normalization constant, $K\in \mathfrak{su}(N)$ is the Cartan subalgebra of $\mathfrak{su}(N)$ algebra and $(N-1)$  real coefficients  $\mu_s(\boldsymbol{\nu})$ are coordinates of   points on a unit sphere $\mathbb{S}_{N-2}(1)$
\begin{equation}
\label{eq:Nsphere}
    \mu_3^2(\boldsymbol{\nu}) + \mu_8^2(\boldsymbol{\nu})
    +\dots+\mu^2_{N^{2}-1}(\boldsymbol{\nu})=1\,.
\end{equation}
In these terms the Wigner function  $W^{(\boldsymbol{\nu})}_{\boldsymbol{\xi}} (\Omega)$ can be represented as 
(see details in \cite{AKhT2018}) 
\begin{equation}
\label{eq:WFsF}
    W^{(\boldsymbol{\nu})}_{\boldsymbol{\xi}} (\Omega) = \frac{1}{N} \left(1 + \frac{N^2-1}{\sqrt{N+1}} (\boldsymbol{\xi}\,, \boldsymbol{n})\right)\,, 
\end{equation}
where  ($N^2-1$)-dimensional vector $\boldsymbol{n} = \mu_3 \boldsymbol{n}^{(3)} + \mu_8 \boldsymbol{n}^{(8)} + \dots + \mu_{N^{2}-1}\boldsymbol{n}^{(N^{2}-1)}$\, is a superposition of $(N-1)$ orthonormal vectors whose components are determined by diagonalyzing  matrix,   
\begin{equation}
 \label{eq:ns}  
   {n}^{(s)}_{\mu} = \frac{1}{2}\,\mbox{tr}\left(
U\lambda_{s}
U^\dagger\lambda_\mu
\right)\,, \qquad  s=3, 8, \dots,  N^2-1\,. 
\end{equation}

\section{KZ-indicator as unitary invariant }
\label{sec:InvKZ}

Now we will discuss  $SU(N)$ invariance of the indicator of nonclassicality  originated from 
the unitary symmetry of a quantum system.

Below it will be argued that the KZ\--indicator is a scalar function which  depends  only on the  SU(N) group  invariants built out of a density matrix $\varrho$ and Stratonovich-Weyl kernel
$\Delta(\Omega_N)$. This statement follows from  the $SU(N)$ covariance properties of states and SW kernels. 
Indeed, according to the covariance axiom (\ref{eq:covSW}),  the rule (\ref{eq:ztransf})  ensures the following  relation:
\begin{equation}
\label{eq:WFtr}
W_{U\varrho U^{-1}}(V^{-1}\Delta\left(z)V\right) =
W_{\varrho}\left(\Delta(T_{{}_{VU}} z )\right)\,,
\end{equation}
for all $U , V  \in SU(N)\,.$  
Now, in order to prove the invariance of the KZ\--indicator, it is convenient to consider the phase space $\Omega_N $ as an embedded subspace, 
$\Omega_N=SU(N)/H \subset SU(N)$\,, with some isotropy subgroup $H\,.$ 
Then, since the Wigner function depends only on the coset coordinates  $\boldsymbol{z} \in \Omega_N$\,, one can extend the integration in (\ref{eq:KZ}) 
to the whole  $SU(N)$ group as  follows: 
\begin{equation}
\label{eq:KZSU(N)}
    \delta\left(\varrho  \,|\, \Delta \right)=Z_N^{-1}\,
\int_{SU(N)}\mathrm{d}\mu_{SU(N)}\,\bigl|W_\varrho\left
(
\Omega_N
\right)\bigl|-1\,.
\end{equation} 
Furthermore, identifying the measure $\mathrm{d}\mu_{SU(N)}$ in (\ref{eq:KZSU(N)}) with the  normalized bi-invariant Haar measure, one can fix the normalization constant,   $Z_N=1/N\,.$ 
Hence, using the property (\ref{eq:WFtr}) and representation (\ref{eq:KZSU(N)}),
one can get convinced that the effect of the $SU(N)$ group action,  $\varrho^\prime= U\varrho U^{-1}$ and  $\Delta^\prime =V^{-1}\Delta V\,,$ leaves the KZ-indicator unchanged,   
\begin{equation}
    \delta\left(\varrho^\prime \,|\,\Delta^\prime \right)=Z_N^{-1}\,
\int_{SU(N)}\mathrm{d}\mu_{SU(N)}\,\bigl|W_\varrho\left
(T_{{}_{VU}} z 
\right)\bigl|-1\,=
     \delta\left(\varrho  \,|\, \Delta \right)\,.
\end{equation}

In the last equality the inverse transformation $z  \to z^\prime =T^{-1}_{{}_{VU}}z$ 
has been performed 
taking  into account the $SU(N)$ invariance of the Haar measure.

Finally, we present two additional  observations on KZ indicator.  
The essence of the indicator reveals itself the best for multypartite systems. Indeed, if one considers such a system described by density matrix Eq.~(\ref{eq:DM}) than  Eq.~(\ref{eq:WFsF}) implies that for $\xi\leq\xi^{*}\leq \xi_{\text {mb}}$, where $\xi^{*}=\frac{1}{(N-1)\sqrt{N+1}}$,  the Wigner function is non-negative. Here, $\xi_{\text mb}$ is the radius of the maximal ball inscribed into the set of mixed states \cite{BenZycz}. It has been proven that all states lying in this ball are absolutely positive partial transpose states and moreover, they are absolutely separable (they can not be entangled by global unitary transformations) \cite{Zyc1, GurBarnum}.  Thus, all the states with Bloch vector less than $\xi^{*}$ are guaranteed to be  absolutely separable.  There is no doubt, that the issue of negativity of the Wigner function for multipartite systems and its relation to quantum correlations is very involved and requires a separate consideration.

On the other hand, an upper boundary for the linear entropy through $\delta(\varrho)$ may be given. 
Immediately from H{\"o}dlers inequality  and the definition of $\delta$ it follows that
\begin{equation}\label{ineq1}
    \delta(\varrho)+1\leq\left(N \int  W_{\varrho}(\Omega_{N})^{2}d\, \mu_{SU(N)}\right)^{\frac{1}{2}}\left(N \int d\, \mu_{SU(N)}\right)^{\frac{1}{2}}\,,
\end{equation}
where the integration is performed with respect to the $SU(N)$ invariant normalized Haar measure. Since the linear entropy  is 
\begin{equation}
    S(\varrho)=1-\mbox{tr}[\varrho^{2}]=1-N \int  W_\varrho(\Omega_{N})^{2}d\, \mu_{SU(N)}\,,
\end{equation}
then the inequality (\ref{ineq1}) provides
\begin{equation}
     S(\varrho)\leq1-\frac{\delta(\varrho)+1}{\sqrt{N}}\,.
\end{equation}

\section{KZ-indicator of a single qubit }
\label{sec:QuantumnesQubit}

For $N=2$ the master equations (\ref{eq:SWMaster}) determine the spectrum of a qubit Stratonovich-Weyl kernel uniquely: 
\begin{equation}
 \mbox{\bf spec}\left(\Delta(\Omega_2)\right) =\{ \frac{1+\sqrt{3}}{2}\,,
\frac{1-\sqrt{3}}{2}\}\,.
\end{equation}
If the unitary factor $U(\Omega_2)$ in SVD decomposition of the Stratonovich-Weyl kernel  is given in 
the symmetric 3-2-3  Euler parameterization: 
\begin{equation}
U(\Omega_2)= 
\exp{\imath\frac{\alpha}{2}\sigma_3}\, 
\exp{\imath \frac{\beta}{2}\sigma_2}\, 
\exp{\imath \frac{\gamma}{2}\sigma_3}\,, \quad  
\end{equation}
with $
\alpha\in [0,2\pi]\,, \;
\beta\in [0,\pi]\,, \;
\gamma\in [0,4\pi]$\,, then the Euler angles $\alpha$ and $\beta$ are  coordinates of  2-dimensional symplectic manifold $\Omega_2= SU(2)/U(1)\,$   and  the Wigner function (\ref{eq:WFsF}) of qubit reads 
\begin{equation}
\label{eq:QubitWF}
    W_{\boldsymbol{\xi}}(\Omega_2)= 
    \frac{1}{2} + \frac{\sqrt{3}}{2}\; (\boldsymbol{\xi}\,, \boldsymbol{n})\,.
\end{equation}
\begin{figure}
\begin{center}
\includegraphics[width= 0.45 \linewidth]{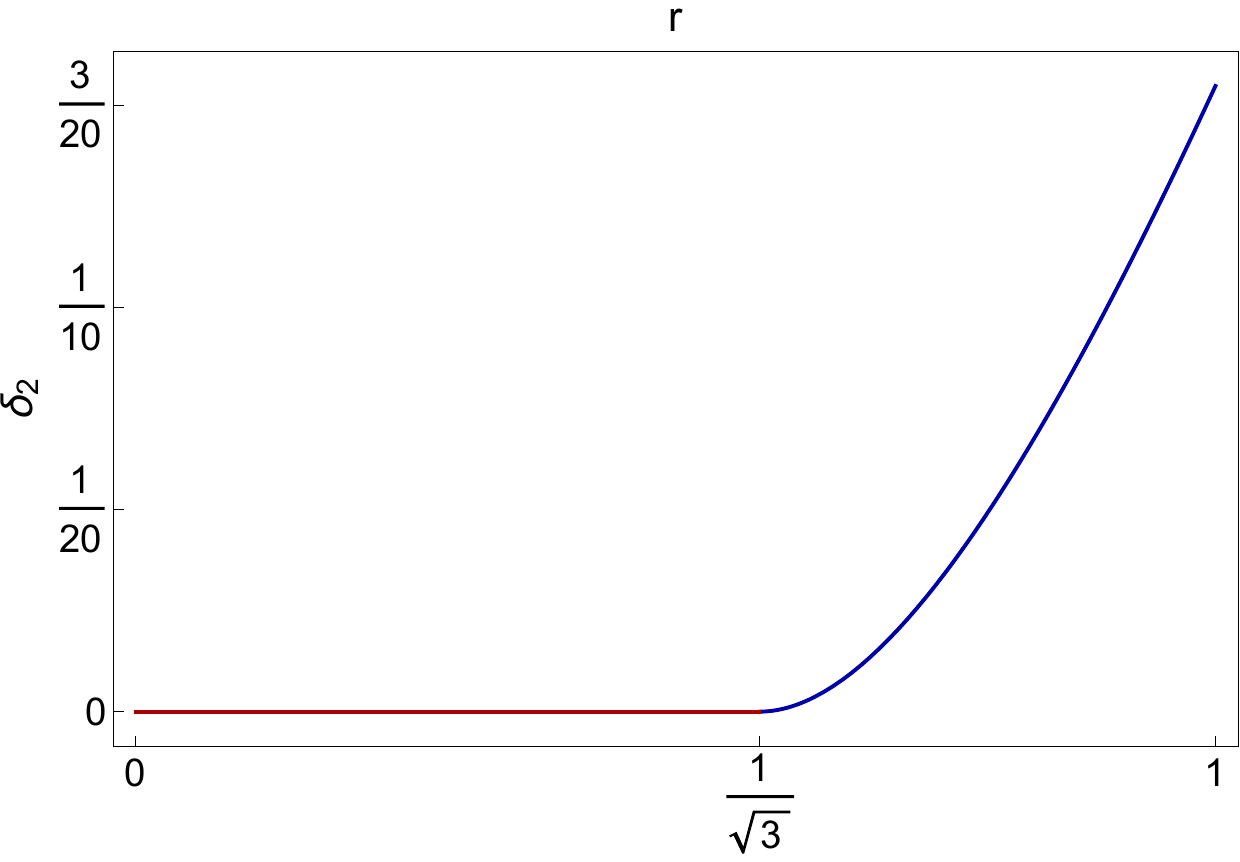}
\end{center}
\caption{\label{fig:QubitKZ}
KZ-indicator for a single qubit  (\ref{eq:qubitKZ}) is zero for the  Bloch radius $r \in [0, 1/\sqrt{3}]$.}
\end{figure}
Here,  the unit vector $\boldsymbol{n}=\left(-\cos\alpha\sin\beta\,, \sin\alpha\sin\beta\,, \cos\beta \right)$\,
parameterizes  $\Omega_2$\,, 
and  $\boldsymbol{\xi} = \left(r\sin\psi\cos\phi\,, \, r\sin\psi\sin\phi\,, \, r\cos\psi\right)$\, is the Bloch vector of qubit in 
a mixed state, 
\begin{equation}
\label{eq:DMqubit}
\varrho =\frac{1}{2}\,\left(\mathbb{I}_2 + \left(\boldsymbol{\xi}, \boldsymbol{\sigma}\right)\right)\,.
\end{equation}
Hence, taking into account that  $\Omega_2 \sim  \mathbb{S}^2(1)\, $ with the standard induced measure, one can write the integral representation for the KZ\--indicator:
\begin{eqnarray}
\label{eq:KZqubitInt}
\delta_2 (r) = 
\frac{2}{2 \pi^2} \frac{1}{2^3}\, 
\int_{0}^{4 \pi} d\gamma 
\int_{0}^{\pi} d\beta
\int_{0}^{2 \pi}
\, d\alpha
\bigg|W_{\boldsymbol{\xi}}(\Omega_2) \bigg| \, \sin{(\beta)}  - 1\,.
\end{eqnarray}
A straightforward evaluation of the integral  (\ref{eq:KZqubitInt}) gives:
\begin{equation}
\label{eq:qubitKZ}
\delta_{2} (r) = 
\left\{\begin{array}{l}
\hspace{1cm}
\  0\,,  \hspace{1.35cm} \qquad\mbox{for}\qquad 
0\leq r \leq  \displaystyle{\frac{1}{\sqrt{3}}}  \,,  \\
\\
\displaystyle{ \frac{\sqrt{3}}{2}\left(r+\displaystyle{\frac{1}{3r}}\right)-1}\,, \qquad\mbox{for}\qquad 
 \displaystyle{\frac{1}{\sqrt{3}}} < r \leq  1\,. 
\end{array}\right.
\end{equation}

\section{KZ-indicator of a single qutrit }
\label{sec:QuantumnesQutrit}

\begin{figure}[hbt]
\begin{minipage}{0.45\textwidth}
\includegraphics[width=\linewidth]{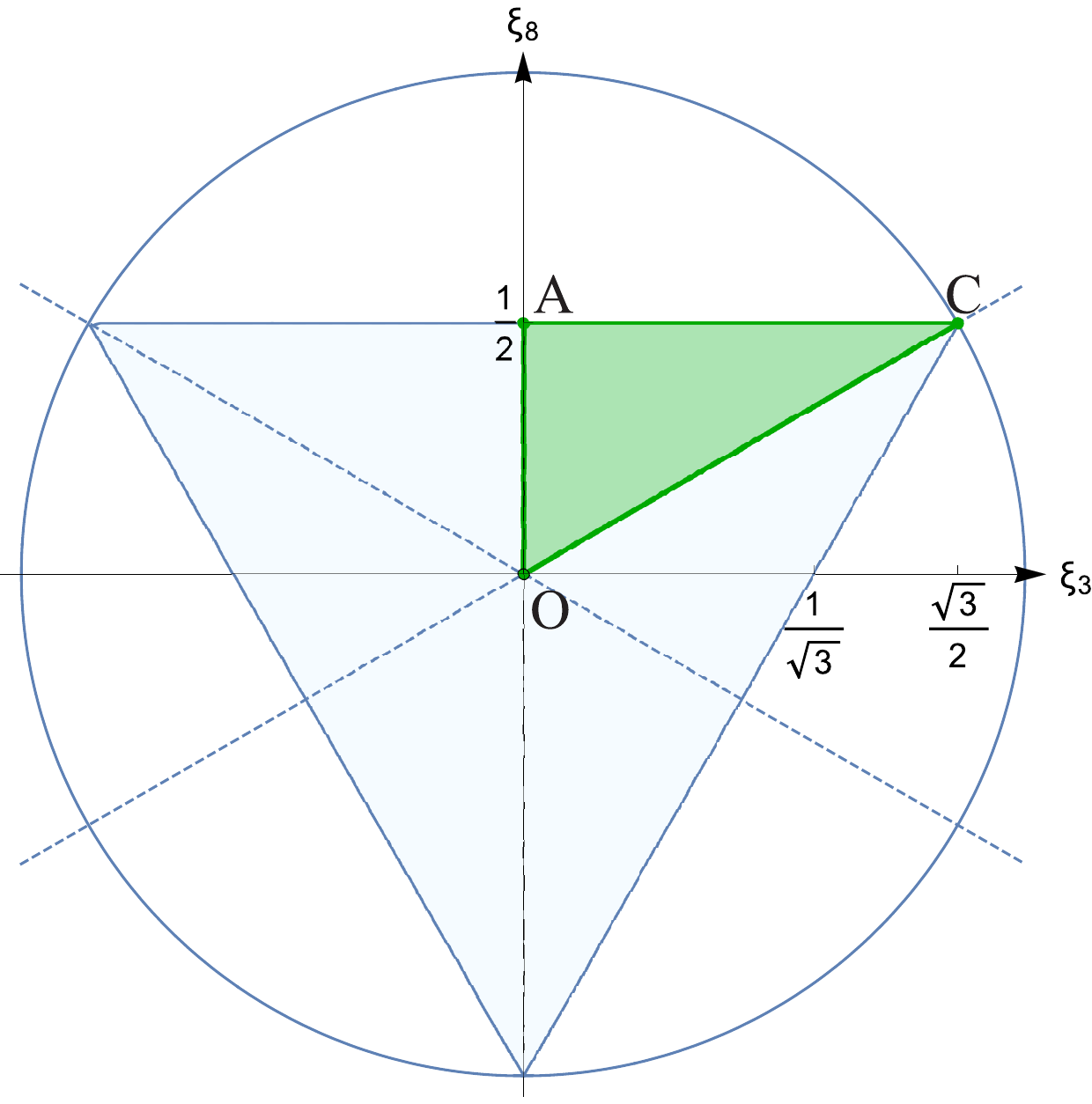}
\caption{
\label{fig:QutritDM-sgn}
Triangle $AOC$ as the  orbit space of qutrit.   \textit{Regular} 6D orbits with $r_1 >r_2>r_3$
correspond to  $\Delta_{AOC} / \{AO\,, OC\}$;  \textit{Degenerate} 4D orbits ($r_1=r_2>r_3$ and  $r_1>r_2=r_3$) correspond to the boundary segments $\{AO\,, OC\}/\{O\}$;  maximally mixed state
$O$ represents the  exceptional orbit.}
\end{minipage}
\hfill
\begin{minipage}{0.45\textwidth}
\includegraphics[width=\linewidth]{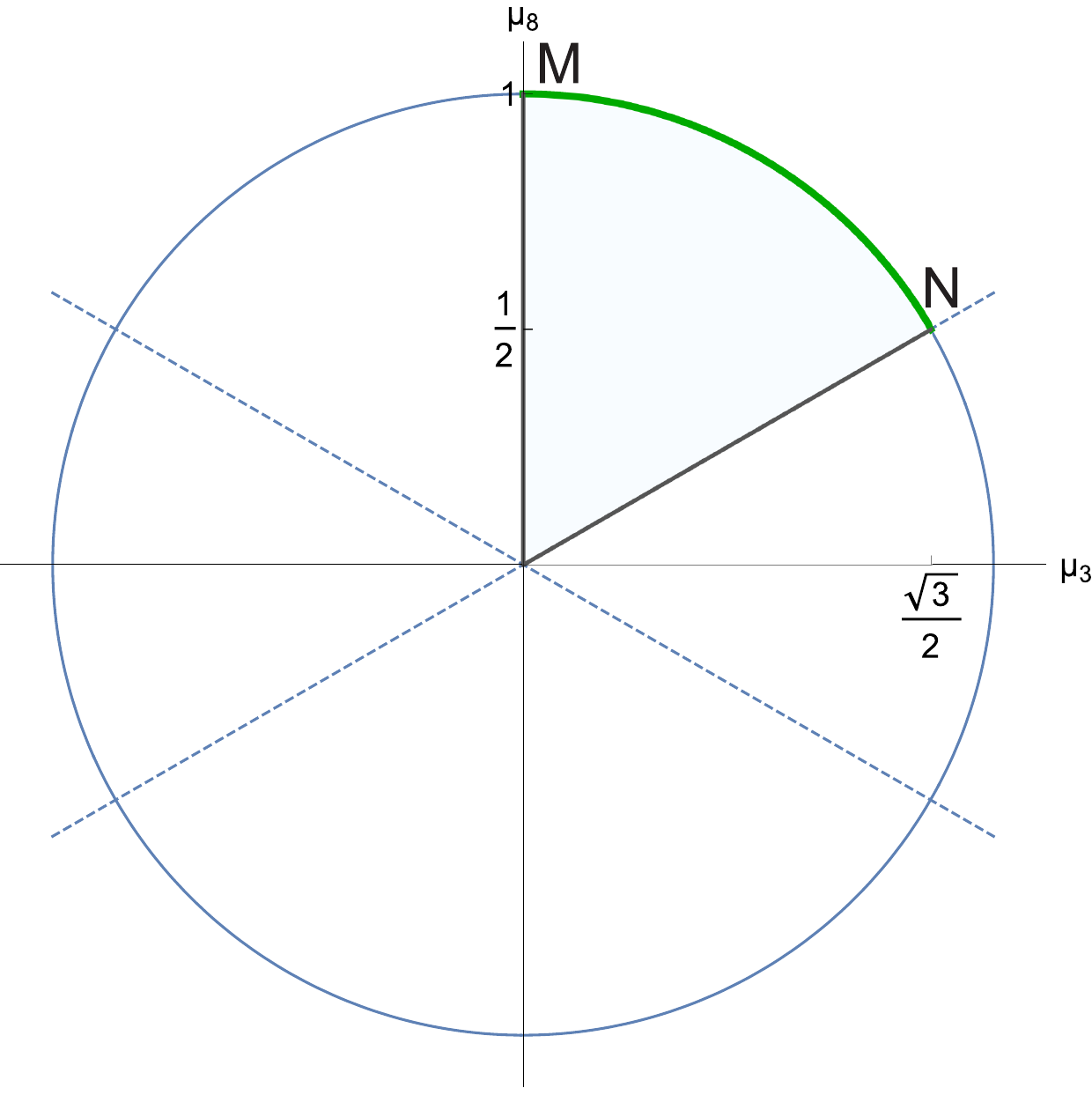}
\caption{
\label{fig:QutritModuliSpace-sgn}
The moduli space of a qutrit Stratonovich-Weyl kernel is given by the arc of a unit circle centered at the origin of $(\mu_3\,, \mu_8)$-plane; it is a union of the 
\textit{regular stratum}, the arc $\wideparen{MN}/\{M\,, N\}$\,, and two \textit{degenerate} Stratonovich-Weyl kernels correspond to the edge points  $M$  and $N$ of the segment.}
\end{minipage} 
\end{figure}

The three-level system in a mixed state $\varrho$ is characterized by the $8$\--dimensional Bloch vector  $\boldsymbol{\xi} = \{\xi_1\,, \dots\,, \xi_8\}$\,:
\begin{equation}
\label{eq:QutritDM}
 \varrho_{3}=\frac{1}{3}\,\mathbb{I}_3 + \frac{1}{\sqrt{3}}\,(\boldsymbol{\xi} , \boldsymbol{\lambda})\, . \end{equation}
In (\ref{eq:QutritDM}) the standard  Gell-Mann basis $\boldsymbol{\lambda} = \{\lambda_1\,, \dots\,, \lambda_8\}$ of $\mathfrak{su}(3)$ algebra is used.
 Based on the  
$SU(3)$ invariance of the KZ\--indicator shown in section \ref{sec:InvKZ}, one can pass to the basis where the density matrix $\varrho$ is diagonal, i.e., the Bloch vector is of the form 
$\boldsymbol{\xi} = \{0, 0, \xi_3, 0, 0, 0, 0, \xi_8\}$,
\begin{equation}
\label{eq:QutritDiagDM}
\varrho=\mbox{diag}||r_1\,, r_2\,, r_3||=\frac{1}{3}\mathbb{I}_3 + \frac{1}{\sqrt{3}}\, (\xi_3\lambda_3 + \xi_8\lambda_8)\,.  
\end{equation}
Below we will assume that the  eigenvalues of a qutrit density matrix belong to the following ordered $C_2$ simplex:
\begin{equation}
\label{eq:Simplex2}
    C_2 = \Big\{ \boldsymbol{r} \in \mathbb{R}^3 \, \Big| \, \sum_{i=1}^{3} r_i = 1,\quad  
    1\geq r_1\geq r_2 \geq r_{3}\geq 0
    \Big\}\,.
\end{equation}
This simplex  represented 
in terms of the Bloch components $\xi_3$ and $\xi_8$\, is given by inequalities  
$$
    0 \leq \xi_3 \leq\frac{\sqrt{3}}{2}\,, \quad 
    \frac{\xi_3}{\sqrt{3}} \leq \xi_8 \leq \frac{1}{2}\,
$$
and is depicted in Fig.\ref{fig:QutritDM-sgn}.

\begin{figure}[hbt]
\begin{minipage}{0.45\textwidth}
\includegraphics[width=\linewidth]{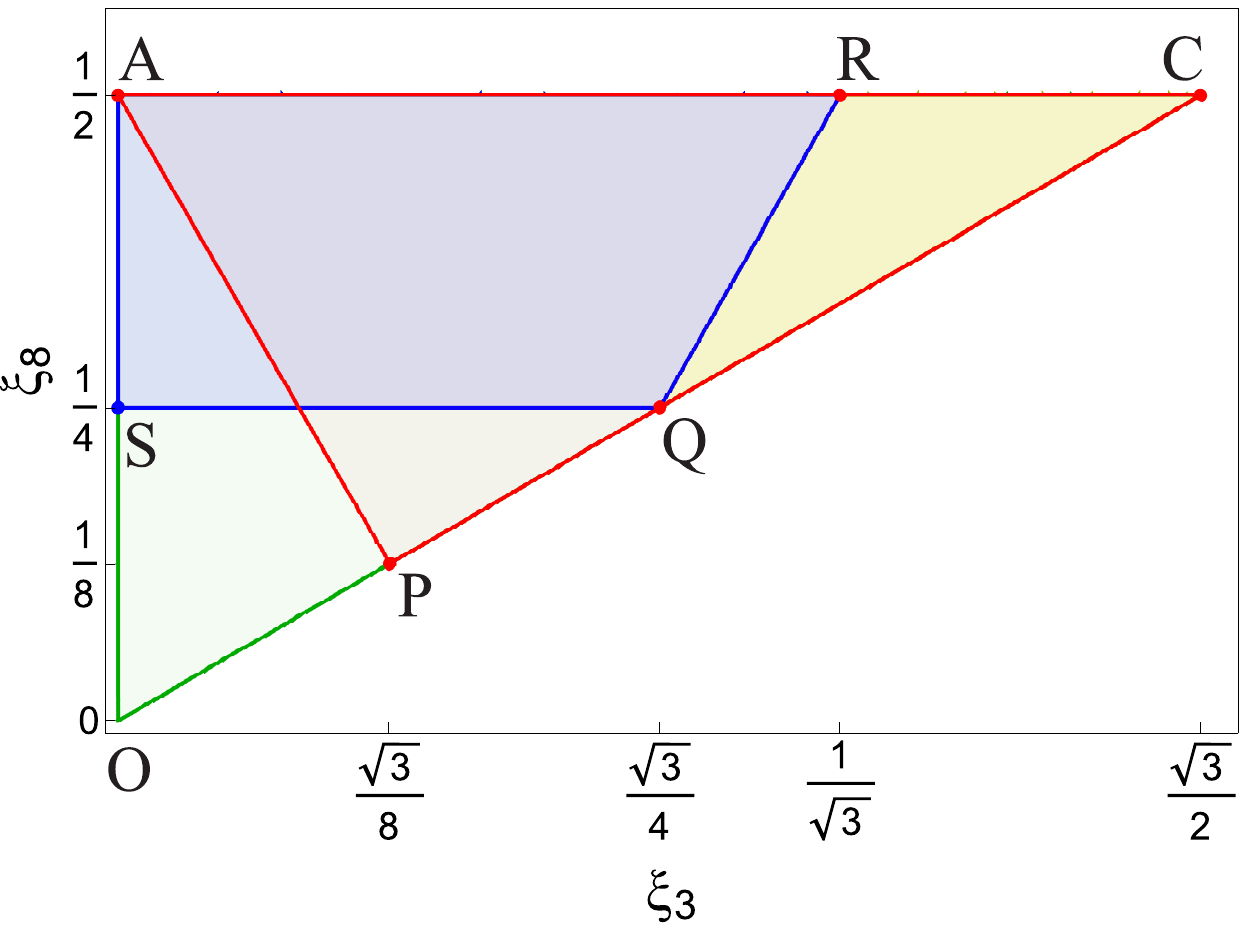}
\caption{
\label{fig:QutritWF-0-pi3-Neg-sgn}
The negativity triangle  $\triangle{APC}$ of a qutrit WF with $\zeta=0\,$. The negativity domain  of a qutrit WF with $\zeta=\frac{\pi}{3}\,$ is the union of trapezium $\square{ARQS}$ and triangle $\triangle{CQR}\,.$}
\end{minipage}
\hfill
\begin{minipage}{0.45\textwidth}
\includegraphics[width=\linewidth]{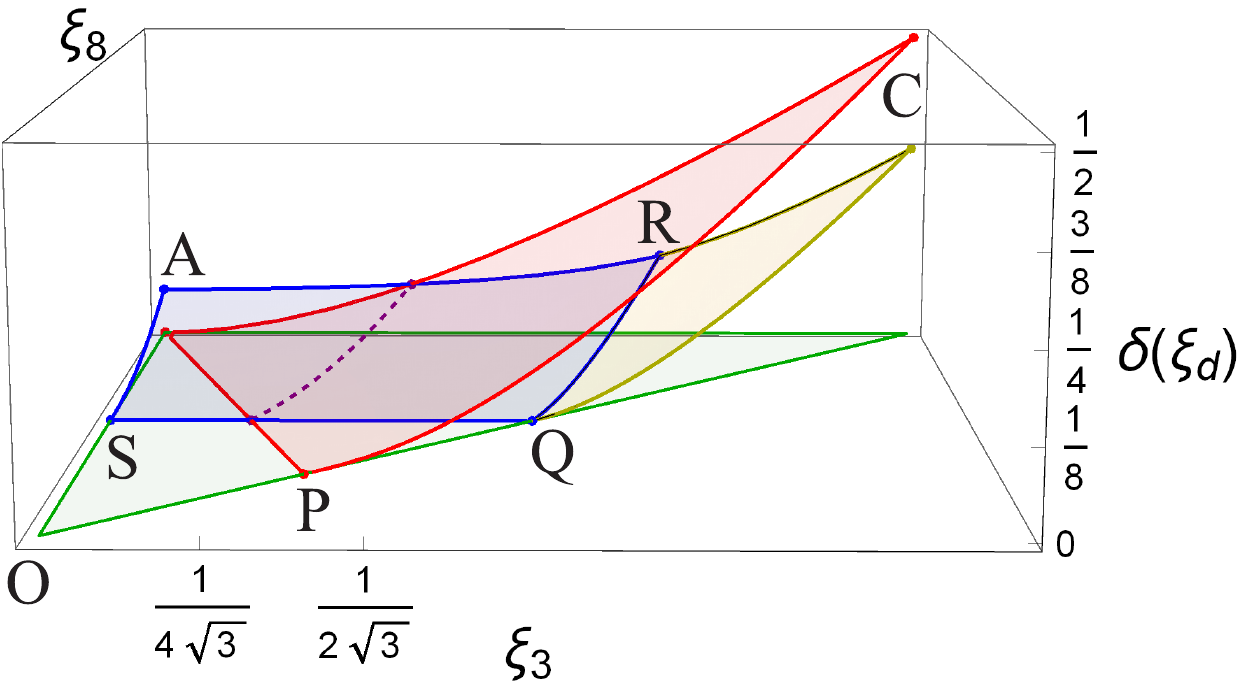}
\caption{\label{fig:QutritKZ-0-pi3-Xi-sgn}
Plots of KZ\--indicators,  $\delta_{1|23}$ (red surface) and $\delta_{12|3}$ (blue and yellow surfaces) of qutrit as functions of two invariants  $\xi_3$ and $\xi_8$\,. 
}
\end{minipage} 
\end{figure}

According to the master equations (\ref{eq:SWMaster}), the spectrum $\mbox{\bf spec}(\Delta(\Omega_3))=\{\pi_1, \pi_2, \pi_3\}$ 
of the Stratonovich-Weyl kernel (\ref{eq:SWk}) of qutrit can be written as
\begin{eqnarray}
\label{eq:specDelta}
\pi_1=\frac{1}{3}+\frac{2}{\sqrt{3}}
\mu_3+\frac{2}{3}\,\mu_8,\quad
\pi_2=\frac{1}{3}-\frac{2}{\sqrt{3}}\mu_3+\frac{2}{3}\,\mu_8, \quad 
\pi_3=\frac{1}{3}-\frac{4}{3}\,\mu_8\,,
\end{eqnarray}
where $\mu_3$ and $\mu_8$ are Cartesian coordinates of a segment of a unit circle (Fig.\ref{fig:QutritModuliSpace-sgn}) with the apex angle $\zeta$:  
\begin{equation}
\label{eq:muzeta}
\mu_3=\sin\zeta\,, \qquad \mu_8=\cos\zeta\,, 
\qquad \zeta \in [0,\  {\pi}/{3}]\,.
\end{equation}
The range of the apex angle corresponds to the  descending order of the eigenvalues, $\pi_1 \geq \pi_2 \geq \pi_3$\,.
The angle $\zeta$ serves as the moduli parameter of the unitary non-equivalent representations of the Wigner functions of qutrit.
Note, that  the edge points M and N of the segment in Fig.\ref{fig:QutritModuliSpace-sgn} with apexes $\zeta=0$ and $\zeta={\pi}/{3}$ 
correspond to degenerate Stratonovich-Weyl kernels:
\footnote{This kernel with the last two equal eigenvalues  was found by Luis \cite{Luis2008}.}
\begin{equation}
\mbox{\bf spec}
\left(\Delta\right)\biggl|_{\zeta=0} =
\left\{1, 1, -1\right\},
\qquad 
\mbox{\bf spec} 
\left(\Delta\right)\biggl|_{ \zeta=\frac{\pi}{3}} =
\frac{1}{3}\left\{5,-1,-1
\right\}\,.
\end{equation}
Depending on the degeneracy of the eigenvalues 
$\pi_1, \pi_2,  \pi_3$\,, we define  the corresponding phase-spaces:
\footnote{Hereafter, following V.I.Arnold  we adopt  notations for ordered set of nonequal eigenvalues by  $(12\dots N)$  and use the sign ``$|$'' between  equal eigenvalues, e.g., $(1|23|4)$.}
\begin{enumerate}
    \item  $\Omega_{(123)}=SU(3)/H_{(123)} $ with the isotropy group
    $H_{(123)}= U(1)^2$
    for a generic Stratonovich-Weyl kernel; 
    \item $\Omega_{(1|23)}=SU(3)/H_{(1|23)}$ with  $H_{(1|23)} \simeq SU(3)/S(U(2)\times U(1))$
    for the Stratonovich-Weyl kernel with the first two equal eigenvalues, $\pi_1=\pi_2$;
    \item $\Omega_{(12|3)}=SU(3)/H_{(12|3)}$ 
    with $H_{(12|3)} \simeq SU(3)/S(U(1)\times U(2))$
    for the Stratonovich-Weyl kernel with the last  two equal eigenvalues, $\pi_2=\pi_3$;
\end{enumerate}
To parameterize all these  factor spaces, we will use the 
generalized Euler decomposition of  $U(\Omega_3)\in SU(3):$ 
\begin{equation}
    \label{eq:EUler3}
U=e^{\imath \frac{\alpha}{2} \lambda_3} 
    e^{\imath \frac{\beta}{2} \lambda_2} 
    e^{\imath \frac{\gamma}{2} \lambda_3} 
    e^{\imath \theta \lambda_5} 
    e^{\imath \frac{a}{2} \lambda_3}  
    e^{\imath \frac{b}{2} \lambda_2} 
    e^{\imath \frac{c}{2} \lambda_3}  
    e^{\imath \phi \lambda_8}\,,
\end{equation}
and the corresponding  normalized Haar measure 
on $SU(3)$: 
\[
\mathrm{d}\mu_{SU(3)} =
\frac{1}{64 \sqrt{3} \pi ^5} \cos{\theta} \sin^3{\theta} \sin{\beta} \sin{b}\,
\mathrm{d}\alpha\wedge\mathrm{d}\beta\wedge\mathrm{d}
  \gamma\wedge\mathrm{d}
  \theta\wedge\mathrm{d}a\wedge\mathrm{d}b
  \wedge\mathrm{d}c\wedge\mathrm{d}
  \phi
\,.
\]
In order to cover ``almost the entire'' $SU(3)$\,, the angles ranges are 
$\alpha,a\in [0,2\pi]$\,, \, $\beta,b\in [0,\pi]$\,, \, $\gamma,c\in [0,4\pi]$\,, \, $\theta\in [0,\pi/2]$\,, \, $\phi\in [0,\sqrt{3}\pi]$\,.  
Gathering  all the above ingredients  together, 
the KZ-indicators for generic and two degenerate Stratonovich-Weyl kernels read 
\begin{eqnarray}
\label{eq:KZQT1}
\delta_{(123)}(\boldsymbol{\xi}_{\mathrm{d}}\,|\,\zeta)&=& \int_{\Omega_{(123)}} \mathrm{d}\Omega_{(123)}\ \biggl|W^{\,(\zeta)}_{\boldsymbol{\xi}_{\mathrm{d}}}(\Omega_{(123)})\biggl| -1\,,\\
\label{eq:KZQT2}
\delta_{(1|23)}(\boldsymbol{\xi}_{\mathrm{d}}\,|\,0)&=& \int_{\Omega_{(1|23)}} \mathrm{d}\Omega_{(1|23)}\ \biggl|W^{\,(0)}_{\boldsymbol{\xi}_{\mathrm{d}}}(\Omega_{(1|23)})\biggl| -1\,,\\
\label{eq:KZQT3}
\delta_{(12|3)}(\boldsymbol{\xi}_{\mathrm{d}}\,|\,\frac{\pi}{3})&=& \int_{\Omega_{(12|3)}} \mathrm{d}\Omega_{(12|3)}\ \biggl|W^{\,(\frac{\pi}{3})}_{\boldsymbol{\xi}_{\mathrm{d}}}(\Omega_{(12|3)})\biggl| -1\,.
\end{eqnarray}
Here the $\zeta$-parametric family of the Wigner function of a qutrit state characterized by the Bloch vector  $\boldsymbol{\xi}_{\mathrm{d}} = \{0, 0, \xi_3, 0, 0, 0, 0, \xi_8\}\,$ is 
\begin{equation}
\label{eq:WFQutritDiag}
 W^{\,(\zeta)}_{\boldsymbol{\xi}_{\mathrm{d}}}(\Omega_{(123)}) =  \frac{1}{3} + \frac{4}{3}\,\left[\,
 \sin(\zeta)\,(\boldsymbol{\xi}_{\mathrm{d}}\,,\boldsymbol{n}^{(3)})
 + 
\cos(\zeta)\, (\boldsymbol{\xi}_{\mathrm{d}}\,,\boldsymbol{n}^{(8)})
 \,
 \right]\,,
\end{equation}
where $
 \boldsymbol{n}^{(3)}$ and $
  \boldsymbol{n}^{(8)}
$ are defined in  (\ref{eq:ns}).
The integration measures in (\ref{eq:KZQT1})-(\ref{eq:KZQT3}) for each stratum is determined by the corresponding isotropy group:
\begin{eqnarray}
\mathrm{d}\Omega_{(123)}&=&
\frac{3\mbox{ Vol}(H_{(123)})}{64\sqrt{3}\pi^5}
\cos{\theta}\sin^3{\theta}\sin{\beta}\sin{b}\,\mathrm{d}
\beta\wedge\mathrm{d}\gamma\wedge\mathrm{d}\theta\wedge\mathrm{d}a\wedge\mathrm{d}b,
\\
\mathrm{d}\Omega_{(1|23)}&=&\frac{3\mbox{Vol}(H_{(1|23)})}{64\sqrt{3}\pi^5}
 \cos{\theta}\sin^3{\theta}\sin{\beta}\sin{b}\,\mathrm{d}\beta\wedge\mathrm{d}\theta\wedge\mathrm{d}b
\,,\\
\mathrm{d}\Omega_{(12|3)}&=&\frac{3\mbox{Vol}(H_{(12|3)})}{64\sqrt{3}\pi^5}
 \cos{\theta}\sin^3{\theta}\sin{\beta}\sin{b}\,\mathrm{d}\beta\wedge\mathrm{d}\theta\wedge\mathrm{d}b\,.
\end{eqnarray}
In order to make our presentation transparent and to simplify the analysis, below we will give only the results of evaluation of the KZ-indicator for two representative Wigner functions whose spectrum is degenerate. 
The evaluation of the integral (\ref{eq:KZQT2}) gives 
\begin{equation}
\label{eq:qutritKZ0}
\delta_{(1|23)}(\boldsymbol{\xi}_{\mathrm{d}})= 
\left\{\begin{array}{l}
\hspace{1cm}
\  0\,,  \hspace{2.2cm} \qquad\mbox{if}\qquad 
\xi_3,\xi_8 \in \triangle  OAP
  \,,  \\
\\
\displaystyle{\frac{1}{36}\frac{(2(\sqrt{3}\xi_3+\xi_8)-1)^3}{\xi_3(\xi_3+\sqrt{3}\xi_8)}}\,, \qquad\mbox{if }\qquad 
\xi_3,\xi_8 \in \triangle  APC\,. 
\end{array}\right.
\end{equation}
Here the triangles $\triangle{OAP}$ and $\triangle{APC}$ decompose simplex $C_2$ in a way shown in 
Fig.\ref{fig:QutritWF-0-pi3-Neg-sgn}. The triangle $\triangle{APC}$ represents the domain of negativity of the Wigner function:
\begin{equation}
\label{eq:QutritBlochWF0Neg}
 \triangle{APC}: =\biggl\{\, 
 \xi_3, \xi_8 \in C_2\ \biggl|\ 
 \frac{1}{8}\leq \xi_8\leq \frac{1}{2}\,, \
 \frac{1-2 \xi_8}{2\sqrt{3}}\leq \xi _3\leq \sqrt{3}\, \xi_8\,
 \biggl\}\,.
\end{equation}
Similarly,  evaluating the integral (\ref{eq:KZQT3}) for the Stratonovich-Weyl kernel with $\zeta=\pi/3$,  we obtain 
\begin{equation}
\label{eq:qutritKZPi3}
\delta_{(12|3)}(\boldsymbol{\xi}_{\mathrm{d}}) = 
\left\{\begin{array}{l}
\hspace{1cm}
\  0\,,  \hspace{2.7cm} \qquad\mbox{if}\qquad 
\xi_3,\xi_8 \in \triangle{OSQ}
  \,,  \\
\\
\ \displaystyle{\frac{1}{18}\,\frac{\left(1-4\xi_8\right)^3}{ \left(\xi_3^2-3\xi_8^2\right)}}\,, \hspace{2.5cm} \mbox{if}\qquad 
\xi_3,\xi_8 \in \square{ARQS}\,, \\
\\
\displaystyle{\frac{1}{36}\,\frac{\left(2(\sqrt{3}\,\xi_3 +\xi_8) +1\right)^3}{\xi_3(\xi_3 +\sqrt{3}\,\xi_8)}-2}\,, \quad\mbox{if }\qquad 
\xi_3,\xi_8 \in \triangle{CQR} \,. 
\end{array}\right.
\end{equation}

The domains of definitions of the both KZ\--indicators as well as their plots are given in Fig.\ref{fig:QutritWF-0-pi3-Neg-sgn} - Fig.\ref{fig:QutritKZ-0-pi3-Xi-sgn} respectively.

\subsection{Qutrit KZ indicator for pure states}

\begin{figure}
    \centering
    \includegraphics[width=0.5 \linewidth]{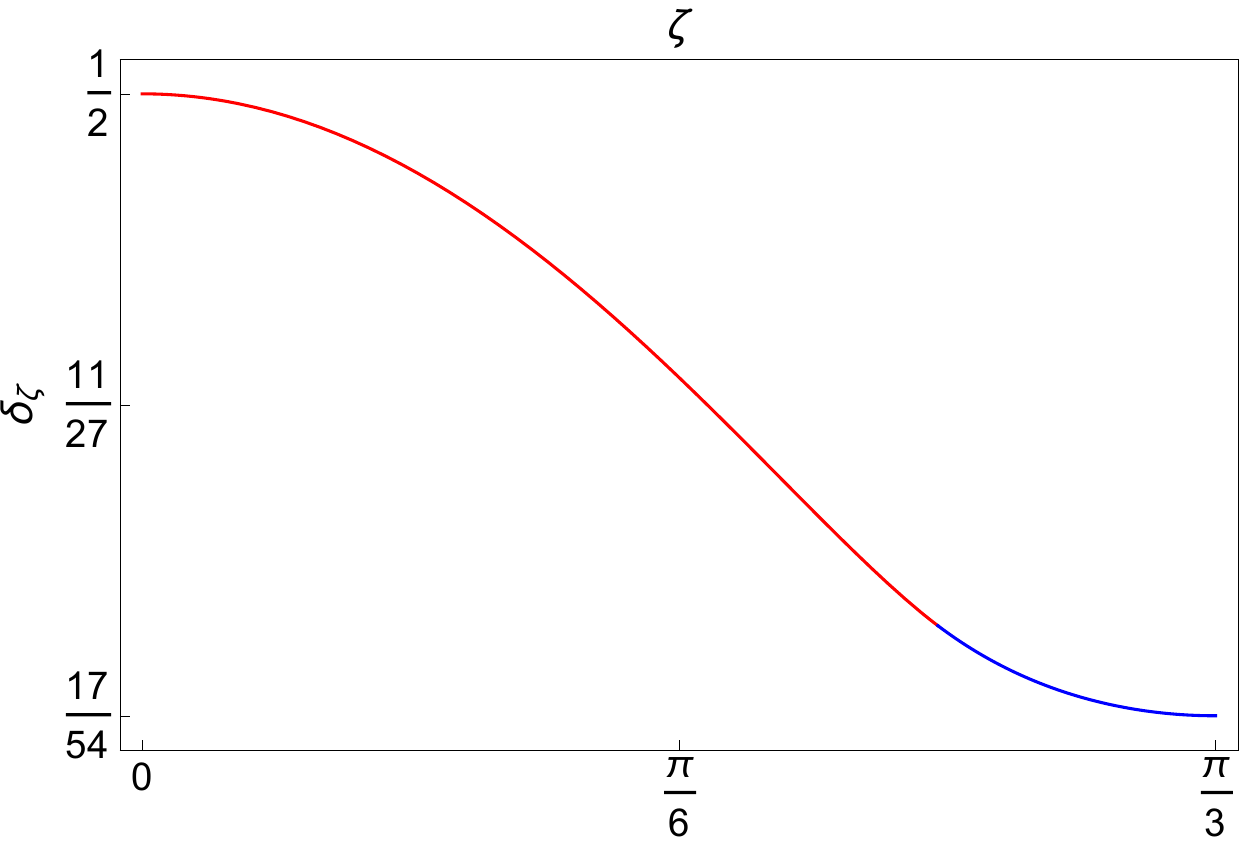}
    \caption{The KZ-indicator as function of moduli parameter for qutrit pure states.}
    \label{fig:KZ qutrit pure}
\end{figure}

In this section we will discuss KZ\--indicator  $\delta_\zeta$ for pure states  of a qutrit, while the Wigner functions  moduli parameter $\zeta \in [0,\, \pi/3]\,$ is arbitrary.
Our calculations show that for all values of the moduli parameter $\zeta$  the indicator 
is  a monotone  decreasing function, $1/2 \leq \delta_\zeta \leq 17/54\,,  $ and it hereby reveals nonclassicality of all pure states (see Fig.\ref{fig:KZ qutrit pure}).

Each pure state of qutrit (i.e. the 
rank-1 state)   belongs to a class of  4-dimensional $SU(3)$ orbits being characterized either by the isotropy group $SU(3)/S(U(1)\times U(2))\,$ or its conjugated group $SU(3)/S(U(2)\times U(1))\,$. 
Let us fix a density matrix $\varrho_0\,$ as the representative of  this class,

\begin{equation}
   \varrho_0=|1\rangle
\langle 1 |=\mbox{\bf diag} \{1\,,0\,,0\,\} =
\frac{1}{3}\,\mathbf{I}_3 + 
\frac{1}{2}\,\lambda_3 + \frac{1}{2\sqrt{3}}\,\lambda_8\,.
\end{equation}
Then an arbitrary pure state $\varrho_{\boldsymbol{\chi}}
=| \boldsymbol{\chi} \rangle
\langle \boldsymbol{\chi} |
$
can be written as 
\begin{equation}
\varrho_{\boldsymbol{\chi}}  =
  V_{\boldsymbol{\chi}} \,| 1\rangle
\langle 1 |\,
V_{\boldsymbol{\chi}}^\dagger\,, \end{equation}
with $V_{\boldsymbol{\chi}}$ from the coset  $SU(3)/S(U(1)\times U(2))$ parameterized by  4 coordinates $\boldsymbol{\chi}$.
Similarly,  according to (\ref{eq:SWk}), a generic SW kernel  is given by the adjoint action of $U_{\boldsymbol{\vartheta}} \in SU(3)/S(U(1)\times U(1)\times U(1))\,$ on a matrix  $P_\zeta
=\mbox{\bf diag}\{\pi_1\,,\pi_2\,,\pi_3\,\}
$:
\begin{equation}
\label{eq:SW}
\Delta({\boldsymbol{\vartheta}}\,|\, \zeta) = U_{\boldsymbol{\vartheta}}\,
\mbox{\bf diag}\{\pi_1\,,\pi_2\,,\pi_3\,\}\,U_{\boldsymbol{\vartheta}}^\dagger \,.
\end{equation} 
In (\ref{eq:SW}) the $6\--$tuple of coordinates of points on $SU(3)/S(U(1)\times U(1)\times U(1))\,$
coset is denoted by 
$\boldsymbol{\vartheta}$\,, and 
the diagonal entries $\{\pi_1\,,\pi_2\,,\pi_3\,\}$ are parameterized according to  (\ref{eq:specDelta}). 
With this input one can get convinced that the
Wigner function  of qutrit in the pure state $ | 
\boldsymbol{\chi}\rangle
\langle \boldsymbol{\chi} |$ 
is  related to the WF of qutrit in the state $| 1\rangle
\langle 1 |$
by the  induced transformation  on phase space, 
\begin{eqnarray}
W_{| \boldsymbol{\chi}\rangle
\langle \boldsymbol{\chi} |}(\vartheta\, |\, \zeta)  
&=&
\mbox{tr}\left(
\varrho_0 V^\dagger_{\boldsymbol{\chi}}
U_{\vartheta}\,P_{\zeta}U^\dagger_{\vartheta}V_{\boldsymbol{\chi}}
\right)
=
\langle 1 |U_{T^{-1}_{\boldsymbol{\chi}}
\boldsymbol{\vartheta}} P_\zeta U^\dagger_{T^{-1}_{\boldsymbol{\chi}}
\boldsymbol{\vartheta}}|1\rangle \nonumber\\
&=&\langle 1 |\Delta (T^{-1}_{\boldsymbol{\chi}}\vartheta\,|\, \zeta)|1\rangle 
=
W_{|1\rangle
\langle 1|}(T^{-1}_{\boldsymbol{\chi}}\vartheta\, |\, \zeta)\,.
\end{eqnarray}
Here, the composition law, 
\(
g^\dagger_{\boldsymbol{\chi}}
g^{}_{\boldsymbol{\vartheta}}= g_{T^{-1}_{\boldsymbol{\chi}}
\boldsymbol{\vartheta}}\,,
\)
and covariance property of SW kernel have been used.

Using this, the computation of KZ-indicator of qutrit gives: 
\begin{equation}
\delta_\zeta=\begin{cases} 
\displaystyle{\frac{(-1+4 \cos (\zeta ))^3}{18 (1+2 \cos (2 \zeta ))}}\,,&\mbox{if}\,\  0\, \leq \zeta \leq 2\arctan\left(\dfrac{\sqrt{3} }{2+\sqrt{5}}\right)\,, \\
    &\\
\displaystyle{
\frac{\left(4 \sin \left(\zeta +\frac{\pi }{6}\right)+1\right)^3}{18 \left(1-2 \cos \left(2 \left(\zeta +\frac{\pi }{6}\right)\right)\right)}}-2\,, &\mbox{if}\,\  2\arctan\left(\dfrac{\sqrt{3} }{2+\sqrt{5}}\right)\leq\zeta\leq\dfrac{\pi}{3}\,. \end{cases}
\label{eq:pure3}
\end{equation}
Note that the expressions (\ref{eq:pure3}) evaluated for the moduli parameters corresponding to the   degenerate SW kernels, $\zeta=0\,$ and $\pi/3 $\,, coincide with the corresponding  limits  of KZ-indicator (\ref{eq:qutritKZ0}) and (\ref{eq:qutritKZPi3}), 
\begin{eqnarray}
\lim_{\xi_3 \to \sqrt{3}/2\,,\, 
\xi_8 \to 1/2}\, \delta_{\zeta=0}(\boldsymbol{\xi}_{\mathrm{d}})=\frac{1}{2}\,,
\quad
\lim_{\xi_3 \to \sqrt{3}/2\,, \,\xi_8 \to 1/2}\, \delta_{\zeta=\pi/3}
(\boldsymbol{\xi}_{\mathrm{d}})
=\frac{17}{54}\,.
\end{eqnarray}

\section{Conclusion}

In the present note we rise the question of dependence of the KZ\--indicators of nonclassicality  on the representation of the Wigner functions. This issue was analysed by constructing the KZ-indicator for two so-called degenerate  Stratonovich-Weyl kernels, which are special representatives of $\zeta$-parametric family of the Wigner function of qutrit. Our calculations show that despite the quantitative  distinction of  these  indicators, there are interesting common features between both indicators of nonclassicality:
\begin{itemize}
\item[\--] If we identify the boundary of a quantum-classical transition as the locus of vanishing Wigner function, then it turns out that quantum-classical transitions are  smooth. Namely, both KZ\--indicators, (\ref{eq:qutritKZ0}) and (\ref{eq:qutritKZPi3}), are smooth  functions on these boundaries. 
\item[\--] The isometries of the state space induce certain symmetry of the nonclassicality indicators.
To find out the roots of this symmetry, note that the triangles $\triangle{OAP}$ and $\triangle{OSQ}$ where  the Wigner function is positive are congruent. By performing rotation of the triangle $\triangle{OSQ}$ on $\pi/3 $ around the point $O$ with subsequent reflection over $OC$\,, one can 
superpose them. This symmetry is a reminiscent of the existence of the Weyl group acting on the eigenvalues of a qutrit density matrix by discrete rotations and reflection. As a result of the Weyl symmetry, one can expect that there are characteristics of the nonclassicality of qubit which are   equal modulo $\pi/3$\,.  From the geometrical reasoning, it is easy to find such characteristics. Indeed, one can get convinced that  for both,   $\zeta=0$
and $\zeta=\pi/3\,,$ the Euclidean areas of the domain where the Wigner function is positive are equal, $S_{\triangle{OAP}}=S_{\triangle{OSQ}} = 2^{-5}\sqrt{3}\,.$ Therefore, assuming that the eigenvalues of qutrit are uniformly distributed, the geometric probability to find a random qutrit state with positive  Wigner function is the same for both degenerate Stratonovich-Weyl kernels with $\zeta=0$ and $\pi/3\,:$
\begin{equation}
\frac{\mbox{\small Euclidean~Area~of~WF~Positive~Part~of~Orbit~Space }}{\mbox{\small Total ~Euclidean~Area~of~Orbit~Space}} = \frac{1}{4}\,. 
\end{equation}
It is clear that the above  argumentation can be  extended to the case of metrics possessing the Weyl symmetry. As an example, one can consider the flat Hilbert-Schmidt metric on a qutrit state space. For this case,  the volume form  on the orbit space reads 
\begin{equation}
\label{eq:qutritOtbitVol}
w_3= \frac{8}{9\sqrt{3}}\,\xi_3^2\left(\frac{\xi_3^2}{3}-\xi_8^2\right)^2\,
   \mathrm{d}\xi_3\wedge \mathrm{d}\xi_8\,,
\end{equation}
and evaluation of  the integrals over ${\triangle{OAP}}\,$ and 
${\triangle{OSQ}}\,$
gives the same results: 
\begin{eqnarray*}
\int_{\triangle{OAP}}\, w_3 
   =\int_{\triangle{OSQ}}\,\, w_3
=\frac{1}{2580480}\,.
\end{eqnarray*}
Hence, noting that 
$$ 
S_{\triangle{OAC}}=\int_{\triangle{OAC}}\, w_3=
\frac{1}{10080}\,,
$$
we conclude that  for both representative WFs the ratio is
\begin{equation}
\label{eq:QR}
\frac{\mbox{\small Hilbert-Schmidt~Area~of~WF~Positive~Part~of~Orbit~Space }}{\mbox{\small Total~Hilbert-Schmidt~Area~of~Orbit~Space}} = \frac{1}{256}\,. 
\end{equation}
\item[\--]Finally, the indicator of nonclassicality  points to the existence of the following three classes of states: 
\begin{itemize}
    \item[$\bullet$] the ``absolutely classical'' states, which have zero KZ\--indicator for all values of the moduli parameters $\zeta$;
    \item[$\bullet$] the ``absolutely quantum'' states, whose KZ\--indicator depends on the moduli parameter but is always non-vanishing;
    \item[$\bullet$] the ``relatively quantum-classical'' states whose classicality/quantumness is  susceptible to a representation of the Wigner function. 
\end{itemize}
Furthermore, all pure states of qutrit belong to ``absolutely quantum'' states.
\end{itemize}

\section{Acknowledgments}
The publication has been prepared  with the support of the “RUDN University Program 5-100” (recipient V.A., KZ-indicator) and of  the grant of plenipotentiary representative of Czech Republic in JINR (symmetry analysis).


\end{document}